\documentclass[showpacs,aps,prl,twocolumn,superscriptaddress]{revtex4}
\usepackage{graphicx} 
\usepackage{dcolumn}
\usepackage{bm}

\begin{document}

\title{Direct Observation of Dark Excitons in Individual Carbon Nanotubes: \\ Role of Local Environments}

\author{Ajit Srivastava}
\affiliation{Department of Electrical and Computer Engineering, Rice University, Houston, Texas 77005, USA}

\author{Han Htoon}
\email[]{htoon@lanl.gov}
\thanks{corresponding author.}
\affiliation{Chemistry Division and Center for Integrated Nanotechnologies, Los Alamos National Laboratory, Los Alamos, New Mexico 87545, USA}

\author{Victor I.~Klimov}
\affiliation{Chemistry Division and Center for Integrated Nanotechnologies, Los Alamos National Laboratory, Los Alamos, New Mexico 87545, USA}

\author{Junichiro Kono}
\email[]{kono@rice.edu}
\thanks{corresponding author.}
\affiliation{Department of Electrical and Computer Engineering, Rice University, Houston, Texas 77005, USA}

\date{\today}

\begin{abstract}
We report the direct observation of the spin-singlet dark excitonic state in individual single-walled carbon nanotubes through low-temperature micro-photoluminescence spectroscopy in magnetic fields.  A magnetic field up to 5~T, applied along the nanotube axis, brightened the dark state, leading to the emergence of a new emission peak.  The peak rapidly grew in intensity with increasing field at the expense of the originally-dominant bright exciton peak and finally became dominant at fields $>$3~T.  This behavior, universally observed for more than 50 nanotubes of different chiralities, can be quantitatively explained through a model incorporating the Aharonov-Bohm effect and intervalley Coulomb mixing, unambiguously proving the existence of dark excitons.  The directly measured dark-bright splitting values were 1-4~meV for tube diameters 1.0-1.3~nm.  Scatter in the splitting value emphasizes the role of the local environment surrounding a nanotube in determining the excitonic fine structure of single-walled carbon nanotubes.

\end{abstract}


\maketitle



\begin{figure}
\includegraphics [scale=0.43] {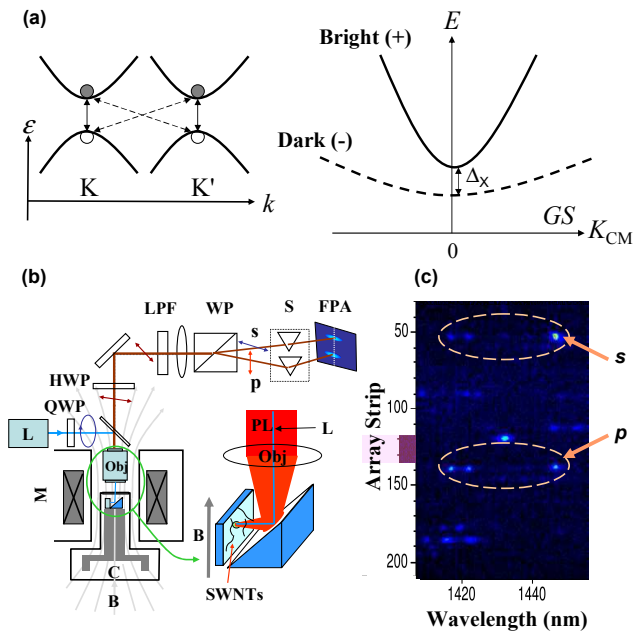}
\caption{(color online).  (a) Schematic energy diagram for spin-singlet excitons in SWNTs.  Left: four possible electron-hole configurations --- direct (KK \& K$'$K$'$, solid) and indirect (KK$'$ \& K$'$K, dashed).  Right: two lowest singlet exciton states --- symmetric ($+$) and anti-symmetric ($-$) combinations of the KK \& K$'$K$'$ direct excitons, ``bright'' and ``dark,'' respectively, with splitting $\Delta_{\rm x}$.  Indirect excitons, not shown, lie at higher energies and are degenerate and optically inactive.  (b) Schematic of the experimental setup used. C: cryostat, M: magnet, L: laser, Obj: objective lens, S: spectrometer, FPA: focal plane array, QWP: quarter-wave plate, HWP: half-wave plate, LPF: long pass filter, and WP: Wollaston prism.  (c) Typical spectra simultaneously showing the $s$ (top) and $p$ (bottom) components of the PL.  The $y$-axis represents the vertical dimension of the 2D array detector.}
\label{figsetup}
\end{figure}

Optical properties of single-walled carbon nanotubes (SWNTs) are affected by their unique band structure together with strong Coulomb interactions characteristic of one-dimensional systems, as shown by recent theoretical~\cite{Ando97JPSJ,SpataruetAl04PRL,ChangetAl04PRL,PerebeinosetAl04PRL} and experimental~\cite{WangetAl05Science,MaultzschetAl05PRB,MortimerNicholas07PRL,ShaveretAl07NL,ShaverKono07LPR,ShaveretAl08PRL} studies.  In semiconducting SWNTs, there are two equivalent valleys (K and K$'$) in momentum space having opposite helicities about the tube axis.  Short-range Coulomb interactions cause inter-valley mixing as well as singlet-triplet splitting, giving rise to exciton fine structure consisting of 
four singlet and twelve triplet, partially-degenerate states.  Among the four singlet states, only one is predicted to be optically active (`bright')~\cite{ZhaoMazumdar04PRL,SpataruetAl05PRL,PerebeinosetAl05NL,Ando06JPSJ}, lying above the lowest energy optically-inactive (`dark') state, as shown in Fig.~\ref{figsetup}(a).  However, there is no consensus as to the value of singlet dark-bright splitting, theoretical predictions ranging from a few to hundreds of meV~\cite{ZhaoMazumdar04PRL,SpataruetAl05PRL,PerebeinosetAl05NL,Ando06JPSJ,Tretiak07NL,JiangetAl07PRB}. A detailed study of the excitonic fine structure of SWNTs is thus essential for resolving these discrepancies and furthering our understanding of radiative and non-radiative energy relaxation processes in SWNTs.

A tube-threading magnetic flux lifts the valley degeneracy and produces two bright excitons due to the Aharonov-Bohm effect~\cite{AjikiAndo93JPSJ,ZaricetAl04Science,ZaricetAl06PRL}.  Previous temperature ($T$)-dependent magneto-optical studies on SWNT ensembles performed at high magnetic fields~\cite{MortimerNicholas07PRL,ShaveretAl07NL,ShaverKono07LPR,ShaveretAl08PRL} have revealed magnetic field ($B$)-induced increase of the photoluminescence (PL) intensity.  However, a direct observation of the dark state was obscured by the broad linewidths of ensemble samples, and the existence of dark excitons was only inferred from the $T$- and $B$-dependences of PL intensity.  The most direct way to ``observe" the dark excitons in SWNTs is to perform spectroscopy on single nanotubes at low $T$ such that the linewidths are smaller than the dark-bright splitting $\Delta_{\rm x}$ and then apply a magnetic field to brighten the dark excitons.  Such methods have been used to prove the existence of dark excitons in semiconductor quantum dots~\cite{BayeretAl00PRB}.  Spectroscopy on a single nanotube can also give us insight into the role of the local environment surrounding the nanotube, which can cause significant changes in emission energies~\cite{HtoonetAl04PRL} and radiative lifetimes~\cite{HagenetAl05PRL}.




Here, we present direct evidence for the existence of the lowest-lying dark exciton state using low-$T$ micro-PL spectroscopy of {\em single} SWNTs in $B$ up to 5~T.  The dark exciton state acquired a finite oscillator strength in a parallel $B$, appearing at a lower energy than the bright exciton peak.  
Due to the very narrow linewidths ($<$1~meV) at low $T$, we could {\em directly} measure the dark-bright splitting without the use of any spectral curve fitting.  Brightening was found to be completely absent when the tube axis was perpendicular to $B$, indicating the crucial role the Aharonov-Bohm phase plays.  A simple model~\cite{ShaveretAl07NL,ShaverKono07LPR,ShaveretAl08PRL} based on Ando's theory of magnetic brightening~\cite{Ando06JPSJ} explains our results remarkably well.  The measured $\Delta_{\rm x}$ values for a number of tubes show scatter, even for tubes with the same chirality.  This clearly suggests the importance of the local environment, particularly local dielectric screening, in determining the excitonic fine structure of SWNTs.  
The values of $\Delta_{\rm x}$ are found to be 1-4~meV for most nanotubes studied. 

We performed magneto-PL spectroscopy of individual micelle-encapsulated HiPco SWNTs deposited on quartz substrates. Figure~\ref{figsetup}(b) schematically depicts the setup.  The sample was cooled in a liquid-helium-flow cryostat in the room temperature bore of a 5~T superconducting magnet.  PL was excited using a tunable Ti:Sapphire laser, collected using a near-infrared objective lens (0.5NA), and detected using a liquid-nitrogen-cooled InGaAs 2D detector array.  To determine the orientation angle $\theta$ of each SWNT with respect to $B$, we relied on their linearly-polarized emission.  A quarter-wave plate was used to circularly polarize the excitation beam to excite all tubes, which were randomly oriented.  PL from SWNTs, linearly polarized along their axes, was sent through a half-wave plate set in such a way that it did not introduce any retardation for the light polarized parallel to $B$.  A Wollaston prism spatially separated the PL into its respective $s$ and $p$ components, which were focused onto different vertical locations along the spectrometer slit.  This enabled us to simultaneously measure the $s$ and $p$ components [as shown in Fig.~\ref{figsetup}(c)] and determine the tube orientation $\theta$.  The substrate was mounted either parallel or perpendicular to $B$.  In the parallel case, due to the random orientation of nanotubes ($\theta$) on the substrate, we could find some nanotubes that were parallel to $B$. However, in the perpendicular case, none of the nanotubes were parallel to $B$.  



\begin{figure}
\includegraphics [scale=0.55]{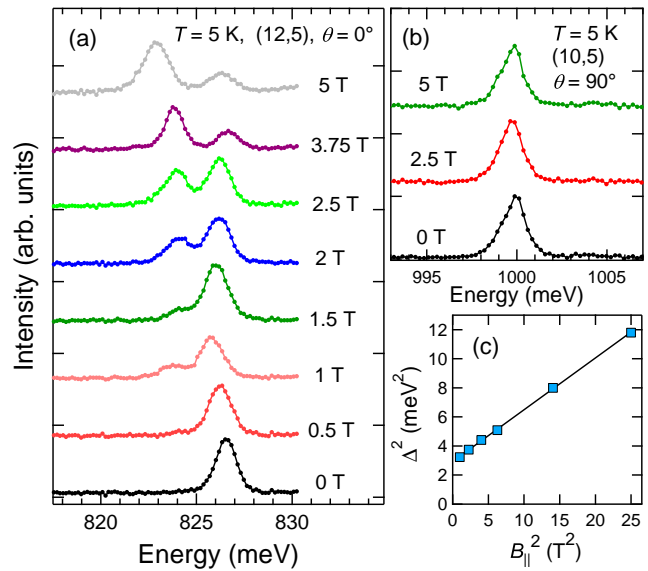}
\caption{(color online).  (a) Typical magnetic-field-dependent PL spectra for a single tube, showing the appearance of a ``dark exciton" peak at a lower energy with respect to the main bright emission peak when a magnetic field is applied parallel to the tube axis. The peak grows with the field at the expense of the bright emission peak and becomes dominant at fields $> 3$~T.  (b) Typical magnetic field dependence for the case of a perpendicular magnetic field.  Clearly, no such magnetic brightening of the dark exciton peak is seen.  (c) Linear fit of the data in (a) using Eq.~(1), to obtain $\Delta_{\rm x}$ from the offset and $\mu$ from the slope (see text for details).}
\label{figBdep}
\end{figure}


Figure \ref{figBdep}(a) shows typical $B$-dependent single-tube PL spectra taken at $T$ = 5~K.  At $B$ = 0, a single, sharp peak exists.  As $B$ is increased, a second peak appears at a lower energy than the first peak.  With further increase of $B$, this second peak gradually grows in intensity and finally dominates the spectrum.  The splitting between the two peaks also increases with $B$.  This $B$-induced appearance of a lower energy
peak was universally observed in all of the 50+ nanotubes of different chiralities which were {\em not} completely perpendicular to $B$.  In Fig.~\ref{figBdep}(b), typical PL spectra for SWNTs oriented completely perpendicular to $B$ are shown.  Clearly, there is no appearance of any lower energy peak even at the highest fields.  We emphasize that we did not observe two PL peaks for nanotubes that were completely perpendicular to the field.  Even in the parallel configuration, only the tubes which had $\theta < 65^\circ$ showed any observable magnetic brightening of the dark state.  Assuming a completely random distribution of $\theta$, this corresponds to $\sim$ 72\% of all the tubes.  Indeed, out of the 75 nanotubes studied in the parallel configuration, 55 tubes (73\%) showed magnetic brightening as is expected from a random distribution of $\theta$.  This also implies that all tubes in our sample emit from the higher energy state at zero field.

We attribute the appearance of the second peak to the $B$-induced brightening of the lowest-energy singlet dark exciton state.  Its absence in $B$ perpendicular to the tube axis convincingly suggests that brightening is induced by the Aharonov-Bohm phase.  In the presence of $B$, the valley degeneracy is lifted due to time-reversal symmetry breaking~\cite{ZaricetAl06PRL}.  As a result, the dark exciton state acquires a finite oscillator strength.  At the highest fields, emission is dominated by the dark state, owing to the increase in splitting, which depopulates the bright state, as well as the increase in the oscillator strength of the dark state at the expense of the bright state oscillator strength.  

To quantitatively explain our observations, we use a two-level model described in Refs.~\onlinecite{ShaveretAl07NL,ShaverKono07LPR,ShaveretAl08PRL}.  We assume that at zero field the dark state is completely dark while the bright state has a relative oscillator strength of 1.  The total dark-bright splitting $\Delta$ in the presence of tube-threading magnetic flux $\phi$ is given by
\begin{equation}
\Delta^2 = \Delta_{\rm x}^2+\Delta_{\rm AB}^2,
\label{eqn1}
\end{equation}
where $\Delta_{\rm x}$ is the zero-field splitting, $\Delta_{\rm AB} = \mu\phi+\Delta_{\rm dis}$, $\mu$ is a coupling constant, $\phi$ = $\pi d^{2}B_{\parallel}/4$, $d$ is the tube diameter, $B_{\parallel} = B \cos\theta$, and $\Delta_{\rm dis}$ is the disorder-induced zero-field splitting.  In the following, we do not include $\Delta_{\rm dis}$, whose existence would imply zero-field brightening of the dark state. Indeed, we only see a single emission peak from the bright state when $B_{\parallel}$ = 0 for any tube, suggesting that the effect of disorder in brightening the dark state is negligible in our sample.
We use Eq.~(1) to fit the splitting value versus $B_\parallel$, obtained directly from the PL spectra at different fields.  The offset and slope of the linear fit between $\Delta^2$ and $B_\parallel^2$ give us $\Delta_{\rm x}$ and $\mu$, respectively, as shown in Fig.~\ref{figBdep}(c).  

\begin{figure}
\includegraphics [scale=0.52] {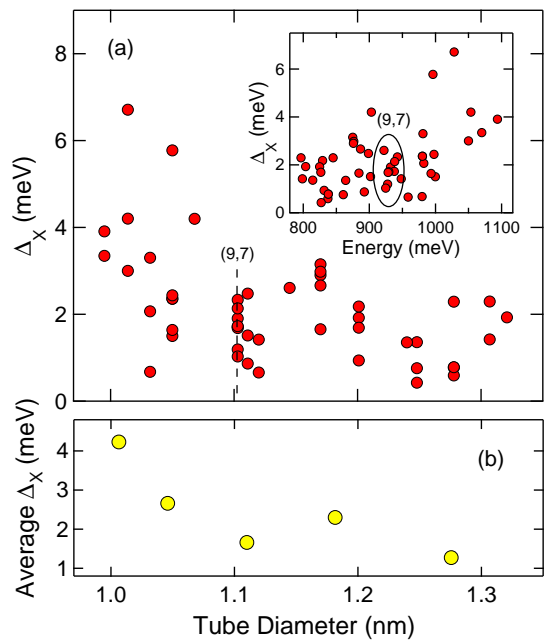}
\caption{(color online).  (a) Measured zero-field dark-bright splitting, $\Delta_{\rm x}$, as a function of tube diameter for 45 tubes.  The data shows scatter even for tubes with the same chirality.  Inset shows the same data plotted against emission wavelength before the assignment of respective chiralities.  Structural assignment was done based on room temperature photoluminescence excitation spectroscopy data taking inhomogeneity of emission wavelength into account, as shown representatively for (9,7) species.  (b) The data in (a) is averaged over 5 diameter ranges and plotted against average diameter to highlight the trend of decreasing $\Delta_{\rm x}$ with increasing diameter.}
\label{figdeltax}
\end{figure}
Figure \ref{figdeltax}(a) shows $\Delta_{\rm x}$ for 45 different tubes versus tube diameter and emission wavelength (inset).  The value of $\Delta_{\rm x}$ is less than 4~meV for the majority of the tubes.  These values (1-4~meV) agree well with those of Refs.~\onlinecite{MortimerNicholas07PRL} and \onlinecite{ShaveretAl07NL} but
should be contrasted to the large (10-100~meV) values predicted by some theoretical studies~\cite{ZhaoMazumdar04PRL,SpataruetAl05PRL,Tretiak07NL,JiangetAl07PRB}.  A possible cause of this discrepancy could be the difference in the value of the dielectric constant ($\epsilon$) used in different calculations.  An accurate prediction must not only take into account the dynamically-screened potential arising from the many-body interactions within the nanotube but also the local variations in $\epsilon$ due to the medium surrounding it (e.g., surfactant, water, quartz substrate).  The dielectric confinement arising from the difference in $\epsilon$ between the nanotube and the surrounding medium causes image charges that can modify both the exciton binding energy and $\Delta_{\rm x}$~\cite{ShabaevEfros04NL,ThoaietAl90PRB}.

Another striking observation one makes from Fig.~\ref{figdeltax}(a) is that the value of $\Delta_{\rm x}$ shows scatter even for tubes of the same chirality.  This scatter must arise from the fluctuations in the local environment of the tubes. Micro-PL spectroscopy is a powerful probe for such fluctuations and enables us to study the role played by the local environment in determining the excitonic fine structure. Since $\Delta_{\rm x}$ is determined by the exchange interaction, it can vary from tube to tube if the local $\epsilon$ varies~\cite{Ando06JPSJ}.  Thus, such local variations of $\epsilon$ can lead to ``inhomogeneous broadening" of $\Delta_{\rm x}$.  For example, according to a recent theoretical estimate~\cite{Perebeinosprivatecomm}, a change of $\epsilon$ = 3 to $\epsilon$ = 2 leads to a change in $\Delta_{\rm x}$ from $\sim$6~meV to $\sim$2~meV for an (11,7) nanotube.
To extract any trend that is present in the $d$ dependence of $\Delta_{\rm x}$, we averaged $\Delta_{\rm x}$ over five $d$ ranges.  This average $\Delta_{\rm x}$, plotted in Fig.~\ref{figdeltax}(b), decreases with $d$, which is reasonable because the short-range Coulomb interaction is predicted to decrease with $d$~\cite{SpataruetAl05PRL,PerebeinosetAl05NL,Ando06JPSJ}.  However, an exact functional form of the $d$ dependence is difficult to deduce from Fig.~\ref{figdeltax}(b) due to the insufficient number of data points.  From Eq.~(\ref{eqn1}) and Fig.~\ref{figdeltax} we obtain values of $\mu$ of 0.4-1.0~meV/T-nm$^2$ for most nanotubes.  These values agree well with the theoretical predictions~\cite{AjikiAndo93JPSJ} as well as previous measurements on ensemble SWNT samples in high magnetic fields~\cite{ZaricetAl04Science,ZaricetAl06PRL,ShaveretAl07NL,ShaverKono07LPR,ShaveretAl08PRL}.  
\begin{figure}
\includegraphics [scale=0.49] {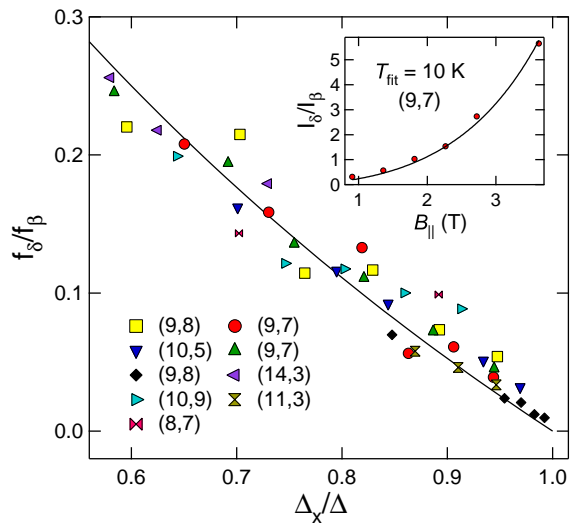}
\caption{(color online).  Dark ($\delta$) / bright ($\beta$) oscillator-strength ratio for 9 different nanotubes versus zero-field dark-bright splitting ($\Delta_{\rm x}$) normalized to the finite-field splitting ($\Delta$).   They all fall onto a single curve given by Eq.~(\ref{eqn3}).  The temperature of all nanotubes was taken to be 10~K, which was obtained by fitting the intensity ratio against $B_{\parallel}$ to Eqs.~(\ref{eqn2}) and (\ref{eqn3}), as shown in the inset.  This value is very close to the measured temperature of 5-7~K.}
\label{fig4}
\end{figure}

Finally, we analyze how the relative intensities of the two peaks change as a function of $B$.  Even though the absolute intensity of each peak was subject to blinking, the relative intensities were not affected by any intensity fluctuations.  Within our two-level model~\cite{ShaveretAl07NL,ShaverKono07LPR,ShaveretAl08PRL}, the intensity ratio of the two peaks is given by
\begin{equation}
\frac{I_{\delta}(B_\parallel)}{I_{\beta}(B_\parallel)} = \exp\left(\frac{\Delta}{k_{\rm B}T}\right)\frac{f_{\delta}(B_\parallel)}{f_{\beta}(B_\parallel)},
\label{eqn2}
\end{equation}
where $k_{\rm B}$ is the Boltzmann constant and $f_{\delta, \beta}(B_\parallel)$ are the oscillator strengths of the dark ($\delta$) and bright ($\beta$) levels in the presence of $B_\parallel$, which vary as
\begin{equation}
\frac{f_{\delta}(B_\parallel)}{f_{\beta}(B_\parallel)} = \frac{1-\frac{\Delta_{\rm x}}{\Delta}}{1+\frac{\Delta_{\rm x}}{\Delta}}.
\label{eqn3}
\end{equation}
Equation (\ref{eqn2}) allows us to extract the oscillator strength ratio $f_{\delta} / f_{\beta}$ from the experimentally-measured intensity ratio ($I_{\delta}/ I_{\beta}$) and splitting ($\Delta$) for each $B_\parallel$.  This quantity, when plotted versus $\Delta_{\rm x}/\Delta$, should exhibit the same functional form given by Eq.~(\ref{eqn3}) for {\em all} nanotubes.  Figure~\ref{fig4} shows the oscillator strength ratio as a function of $\Delta_{\rm x}/\Delta$ for 9 different nanotubes of varying chiralities.  We find that the data points for all nanotubes indeed follow the universal curve given by the right-hand side of Eq.~(\ref{eqn3}), in remarkable agreement with the theory.

In conclusion, we have unambiguously demonstrated the existence of the theoretically-predicted spin-singlet dark exciton state {\em below} the bright state.  The dark-bright splitting was directly measured and found to be 1-4~meV for tubes with diameters of 1.0-1.3~nm.  Scatter in splitting value suggests the importance of the local environment in determining the excitonic fine structure.

\begin{acknowledgments}
We thank the US Army Research Office (No.~49735-PH) and the Robert A.~Welch Foundation (No.~C-1509) for support and V.~Perebeinos for helpful discussions.  Magneto-PL studies reported here were conducted at the Center for Integrated Nanotechnologies, jointly operated for US Department of Energy by Los Alamos and Sandia National Laboratories.
\end{acknowledgments}


\end{document}